\documentclass[11pt]{article}
\usepackage[letterpaper, total={7.5in, 10in}]{geometry}
\usepackage{graphicx}
\usepackage{titlesec}
\usepackage{booktabs}
\usepackage{caption}
\usepackage{listings}
\usepackage{verbatim}
\usepackage{float}

\usepackage{amsmath}
\title{The Performance of MC X-ray and PENELOPE in Homogeneous Bulk Samples}
\author{Dawei Gao, Yu Yuan, Nicolas Brodusch, Raynald Gauvin \\
Department of Mining and Materials Engineering, \\
McGill University, \\
3610 University, \\
Montréal, Québec, Canada, H3A 0C5}

\usepackage{indentfirst}
\begin{document}
\maketitle

\begin{abstract}
This manuscript presents a comparative analysis of two software packages, MC X-ray and PENELOPE, focusing on their accuracy and efficiency in simulating k-ratios for binary compounds and comparing their spectra with experimental data for pure elements and compounds. Based on the Pouchou database, MC X-ray slightly outperforms PENELOPE in k-ratio calculations, achieving a root mean square error (RMSE) of 2.71\% compared to 2.87\%. Discrepancies between the two programs emerge at lower beam energies (3 keV and 5 keV) when comparing simulated spectra with experimental data; however, at higher energies (20 keV and 30 keV), both software packages exhibit consistent and reliable performance across a range of atomic numbers. While both tools are effective for analyzing homogeneous bulk samples, MC X-ray offers significant advantages in processing speed and user-friendliness. This study underscores the strengths and limitations of each package, providing valuable insights for researchers engaged in X-ray simulation and microanalysis.
\end{abstract}

\section{Introduction}

X-ray quantitative analysis is a method for determining the chemical composition of a sample by measuring the intensity of X-rays that are emitted by the sample under electron beam irradiation. This technique is widely used in materials science, chemistry, and geology (Castaing, R. 1951). However, its main drawback is that it shows bad performances with heterogeneous samples because of the complicated relationship between the X-ray emission and internal microstructure of the sample. To overcome this issue, a possible path is to use Monte Carlo modeling to simulate the complex X-ray emission in samples of known internal microstructure and thus address some of the drawbacks of the X-ray quantitative analysis in heterogeneous samples (Gauvin et al., 1992). 

The Monte Carlo method, developed in the mid-1940s (Sood, A 2017), is particularly valuable for tackling problems involving multiple variables. Monte Carlo simulations use random numbers to generate a large number of samples, which helps estimate the governing laws of the system being studied. This method has a well-established history in quantitative analysis, beginning with its application to experimental data on electron scattering in a 1000-angstrom film (Green, M., 1963). Later, Bishop (Bishop, H. E. 1965) was able to derive the  electron backscatter coefficient.

The Monte Carlo method effectively simulates electron-solid interactions by modeling the trajectory of a single electron incident on the sample and repeating the process with different random numbers. The resulting trajectories are displayed in Figure 1 for 1000 electrons and an accelerating voltage of 20 kV. In this figure, the red lines represent the trajectories of the backscattered electrons while the blue lines account for the absorbed electrons, i.e. those who have transferred all their kinetic energy to the specimen. By utilizing random numbers to emulate process variability, Monte Carlo simulations can quantify the number of generated X-rays along the path of each electron in the matter and produce the corresponding X-ray spectrum (Gauvin et al., 2006) after multiple repetitions. 

\begin{figure}[htbp]
\centering
\includegraphics[width=0.6\textwidth]{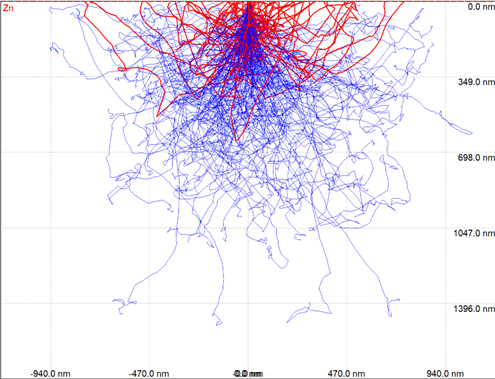}
\caption{Electron trajectories of 1000 electrons at 20 keV in bulk Zn using CASINO 2 (Drouin et al., 1997). The red lines represent the trajectories of the backscattered electrons while the blue lines account for the absorbed electrons}
\end{figure}

To carry out Monte Carlo modelling of electrons in the matter, there are two main programs available in the community. The first one, PENELOPE (Llovet, X., \& Salvat, F. 2006) is capable of modeling the coupled transport of electrons and photons. It tracks primary electrons as well as all secondary (and higher order) electrons and X-rays generated during each interaction of the simulation process. While this comprehensive approach results in a more sophisticated simulation, it also necessitates significantly longer simulation times. On the contrary, another approach to compute the total number of X-rays generated is to compute the number of X-rays generated between two consecutive elastic events and integrate the generated x-ray depth distribution curve (the so-called $\phi \rho z$ curve) reduced by the absorption attenuation factor to calculate the emitted number of X-rays for each element of the sample. Several programs use this approach, the most known being Casino (Drouin et al., 1997), Win X-ray (Gauvin et al., 2006), MC X-ray (Gauvin et al., 2009) and DTSA-II (Ritchie, N. W. 2009). 

MC X-ray is an evolution of Win X-ray (Gauvin et al., 2006) and CASINO (Drouin et al., 1997). While Win X-Ray can only simulate x-ray spectra of homogeneous materials and CASINO performs only the computation of net x-ray intensities in a limited set of geometries, MC X-ray can generate the complete x-ray spectra of sample with various geometries (Gauvin et al., 2009). Recently, the MC X-ray code has been integrated into Dragonfly (ORS, Montreal, Canada), a software platform used for processing scientific 3D image datasets and this integration makes MC X-ray faster and more user friendly (Rudinsky et al., 2019). 

The DTSA-II software (Ritchie, N.W., 2009) provides precise control over detector orientation and enables efficient simulations independent of the solid angle. It also accurately accounts for X-ray absorption and scattering events. Furthermore, the software is capable of calculating secondary fluorescence during electron-solid interactions. Since DTSA-II employs a continuous slowing-down approximation, similar to that used in MC X-ray, Casino, and Win X-ray for calculating X-ray emission, we selected MC X-ray and PENELOPE for performance comparison. The objective of this paper is to assess the accuracy of MC X-ray and PENELOPE by comparing their simulation results under identical conditions on bulk specimens.

\section{Method}
To assess the MC X-ray and PENELOPE programs objectively and thoroughly, a side-by-side evaluation was executed using several criteria:  
\subsection{Comparison of simulated k-ratios with the Pouchou database}
The k-ratio method was first proposed by Castaing (Castaing, R. 1951), it is the ratio of the measured characteristic X-ray intensity of the target element in an unknown specimen to that in a standard of known composition. The k-ratio method is the most comment techniques of quantitative X-ray microanalysis.  The Pouchou database (Pouchou et al., 1991), a portion of which is shown in Table 1 in the Appendix, contains a comprehensive collection of 826 conductive binary compound specimens with well-defined nominal compositions. Each specimen's k-ratio has been experimentally measured, providing valuable data for X-ray analysis. The database encompasses sets of measurements conducted at variable acceleration voltages, ranging from 2.5 kV to 48.5 kV, allowing for a wide range of analysis possibilities. Additionally, the measurements include the K, L, and M series lines of the generated X-rays, enabling detailed investigations into each characteristic lines’ information. Thus, the Pouchou database serves as a valuable database to test the accuracy of simulated k-ratios. 
\subsection{Comparison of simulated spectra with the real experiment data}
To assess the accuracy of the simulated spectra generated by PENELOPE and MC X-ray, experimental spectra were recorded from pure element and compound standards and compared with the simulations. The experimental spectra were acquired using a Hitachi SU-3500 Variable Pressure Scanning Electron Microscope (VP-SEM), equipped with an 80 mm² X-Max silicon drift detector (Oxford Instruments). A list of the standard samples used for data collection is provided in Table 2.

In the simulation step, it was essential to replicate the conditions of the real experiment. Key microscope parameters—including working distance (10 mm), beam current (1 nA), take-off angle (35°), detector radius (5 mm), and beam energy—were matched to the experimental settings to the extent that these parameters were known. For the detector parameters, we ensured the simulation employed the same polar angles and radius as in the actual experiments. The spectrum comparisons were based on 10,000 electrons. Additionally, the simulation parameters for PENELOPE were carefully aligned with those of the real experiment.

The RMSE (Root Mean Square Error) defined by equation (1) is used to quantify the differences between the spectra obtained from MC X-ray simulations and the real experimental data. In this equation, $\hat{y}_i$is the simulated value of a single point (channel of the spectrum) from the MC X-ray simulation and $y_i$ is the simulated value of a single point (channel of the spectrum) from the experimental value of the real experiment, n is the total points (channels) on each spectrum.

\begin{equation}
\text{RMSE} = \sqrt{\frac{\sum_{i=1}^{n} (\hat{y}_i - y_i)^2}{n}}
\end{equation}
\section{Results}
\subsection{Comparison of the k-ratio with Pouchou database}
To compare the k-ratio simulated by MC X-ray and PENELOPE with the Pouchou database we calculated the ratio between the k-ratio from the simulation and the k-ratio from the Pouchou database. A value close to 1 indicates strong agreement between the simulation and experimental results. In contrast, a significant deviation from 1 suggests a lower level of agreement. The comparison results are presented in Figure 2.

Upon analyzing the results of the two cases, we observed that both the ratio of the simulated k-ratio to the Pouchou k-ratio concentrated around 1 for both MC X-ray and PENELOPE. This suggests that both software programs can generate reliable simulation results. Additionally, we found that MC X-ray had a slightly higher degree of precision as its RMSE was only 2.71\% compared to 2.87\% for PENELOPE. And MC X-ray is more concentrated around 1, as 99\% of the MC X-ray data is between 0.9 and 1.1, while 95\% of the PENELOPE data is within this range.

\begin{figure}[htbp]
\centering
\includegraphics[width=0.8\textwidth]{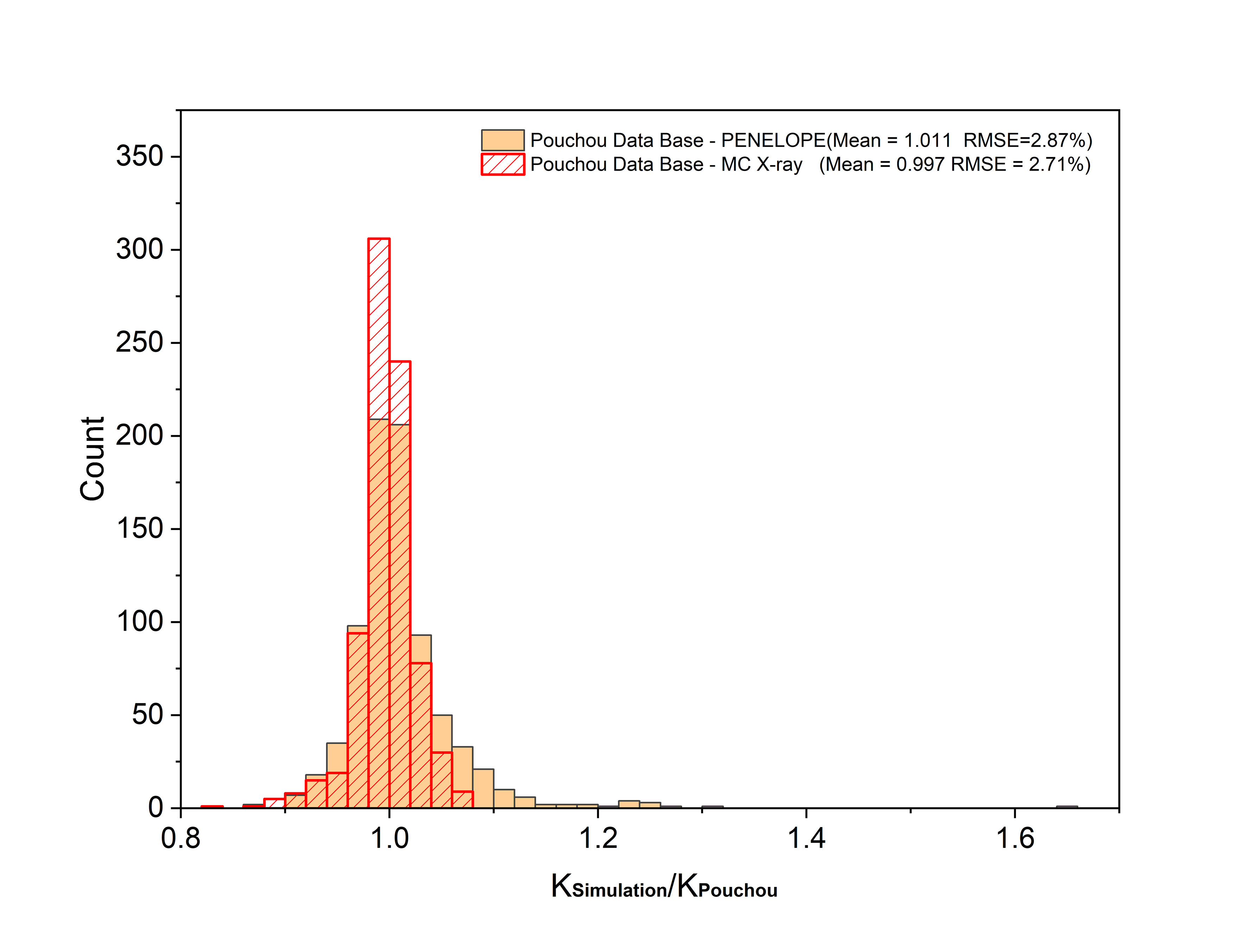}
\caption{The k-ratio comparison results between MC X-ray and PENELOPE using the Pouchou database. The analysis was conducted on 826 conductive binary compound specimens with acceleration voltages ranging from 2.5 kV to 48.5 kV. The data from MC X-ray exhibits a more concentrated distribution}
\end{figure}

\clearpage
\subsection{Comparison of MC X-Ray and PENELOPE simulated spectrum with real experimental data}
To further confirm the accuracy of the simulation results from MC X-ray and PENELOPE, we conducted a comparison between simulated and experimental spectra. 

To test the impact of the atomic number and beam energies on the comparison results, samples with atomic numbers varying from 6 to 83 with 6 different beam energies (3, 5, 10, 15, 20, and 30 keV) were used. A list of the 27 samples analyzed is given in Table 2.  3 representative samples Si (Z=15), Cu (Z=29), and Pt (Z=70) at 3 keV, 5 keV, 20 keV and 30 keV were chosen to show, and then compared with real experimental data. 

The comparison results for Si, Cu, and Pt at 3 keV, 5 keV, 20 keV, and 30 keV beam energies are presented in Figure 3. To enhance the precision of the comparison, we normalized the spectra by dividing each point on the spectrum by the total counts of the entire spectrum and displayed the y-axis on a logarithmic scale.

As shown in Figure 3, there is a significant discrepancy between the simulated and experimental spectra at lower beam energies, such as 3 keV and 5 keV. However, at higher beam energies, such as 20 keV and 30 keV, the simulated and experimental spectra align much more closely. Additionally, it can be observed that at high beam energy (e.g., 30 keV), variations in atomic number have little effect on the differences between simulated and experimental results. In contrast, at lower beam energies (e.g., 3 keV), the discrepancies become more pronounced, especially for samples with higher atomic numbers.

\textbf{3 keV}
\begin{figure}[htbp]
  \centering
  \begin{minipage}[b]{0.3\textwidth}
    \includegraphics[width=\textwidth]{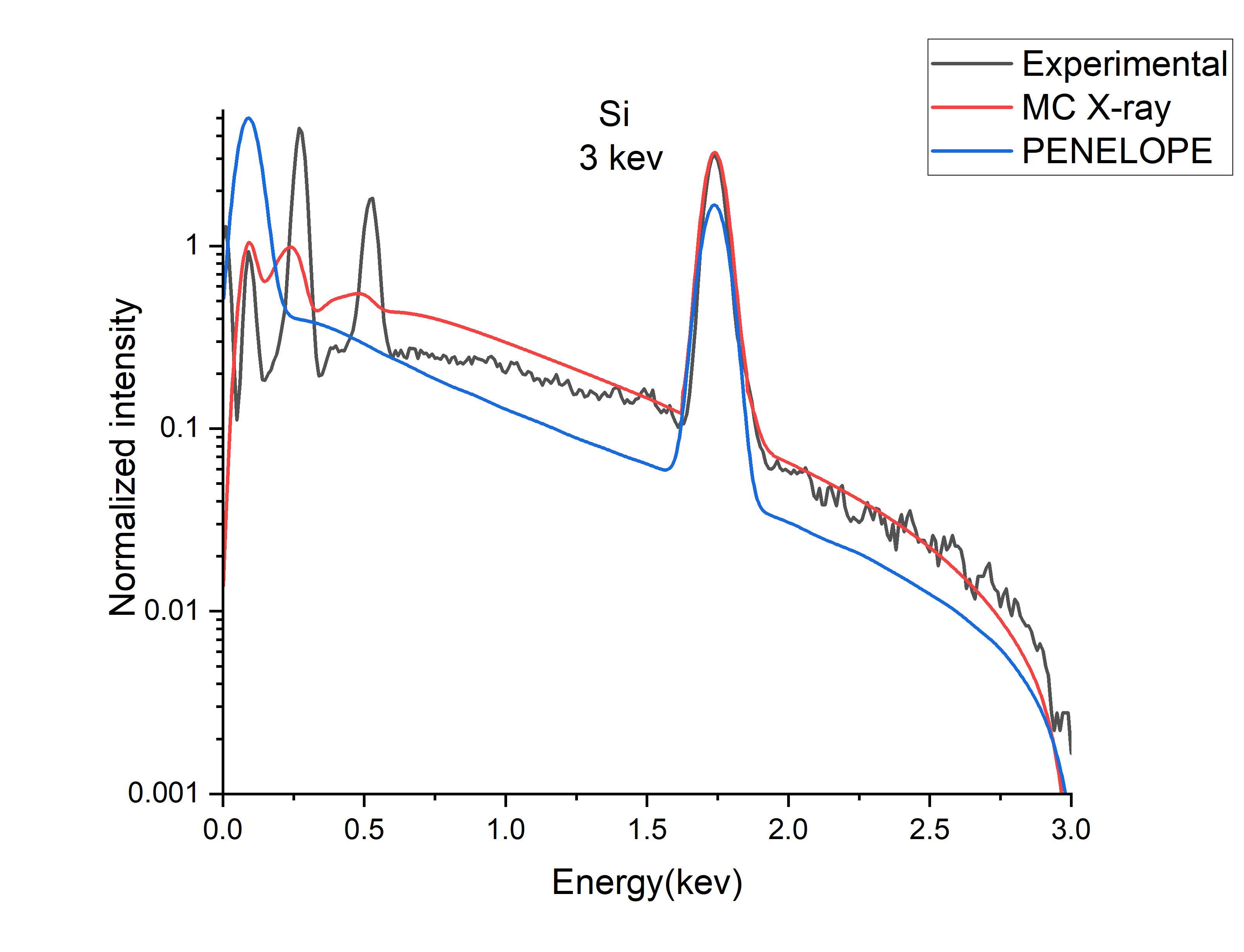}
  \end{minipage}
  \hfill
  \begin{minipage}[b]{0.3\textwidth}
    \includegraphics[width=\textwidth]{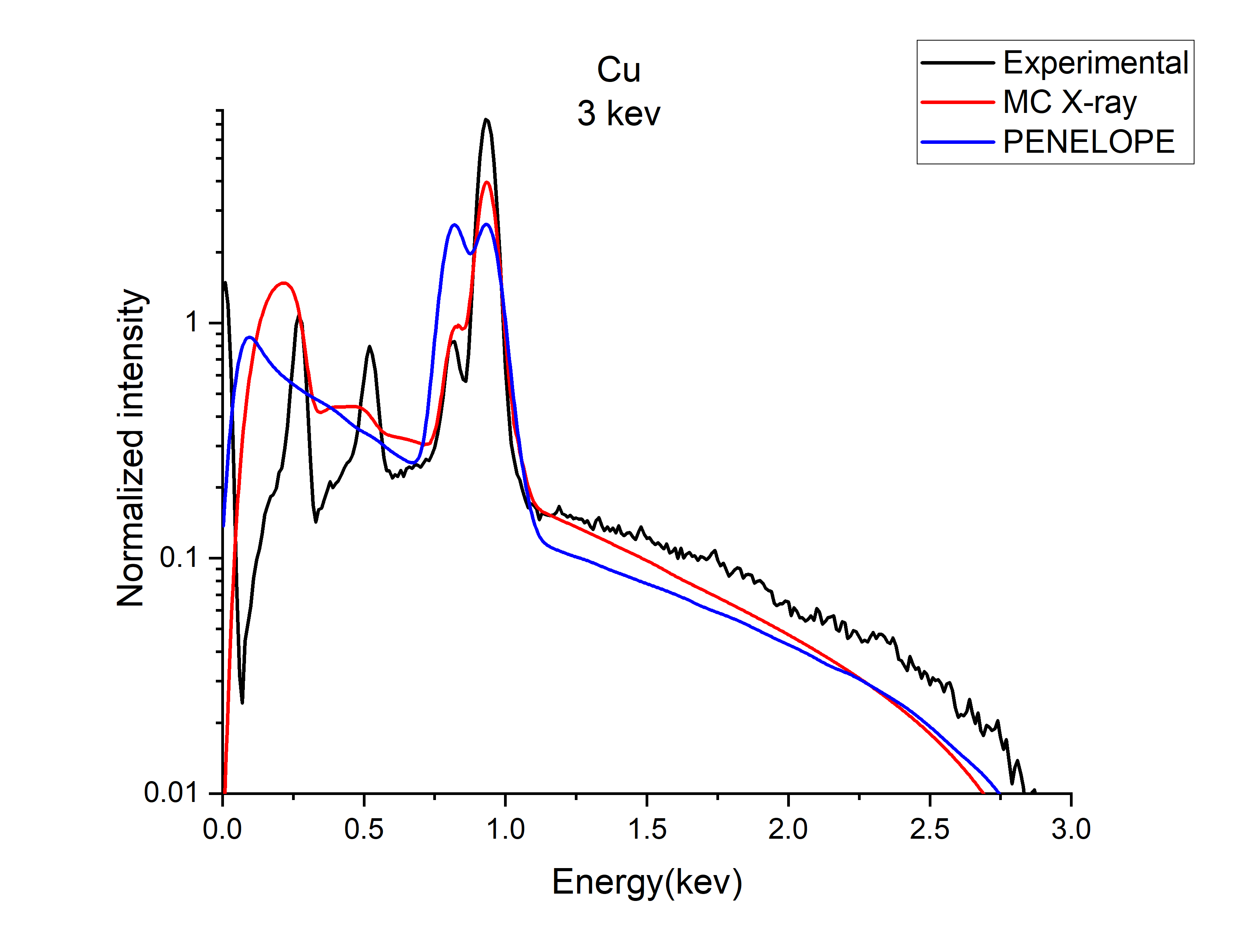}
  \end{minipage}
  \hfill
  \begin{minipage}[b]{0.3\textwidth}
    \includegraphics[width=\textwidth]{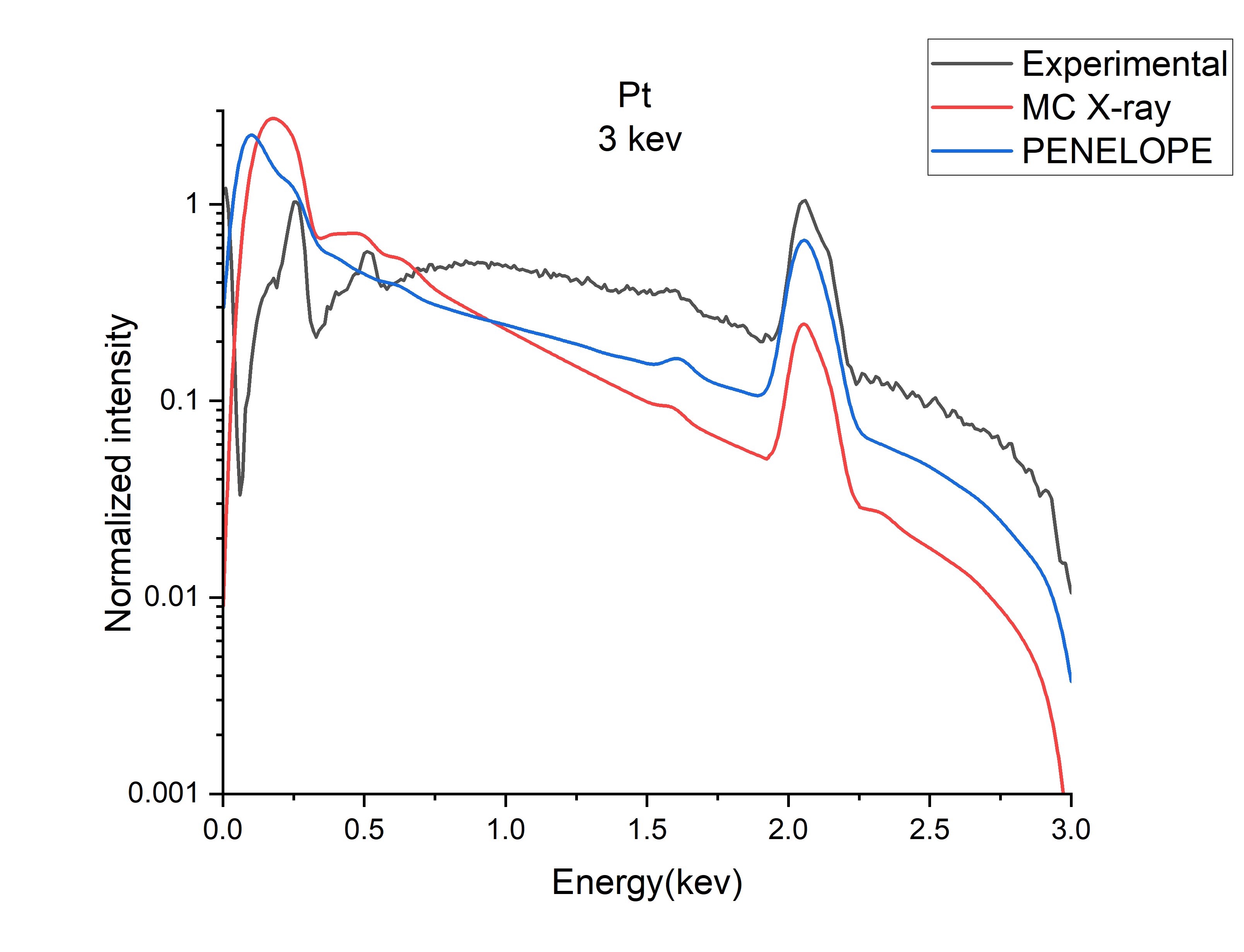}
  \end{minipage}
\end{figure}

\textbf{5 keV}
\begin{figure}[htbp]
  \centering
  \begin{minipage}[b]{0.3\textwidth}
    \includegraphics[width=\textwidth]{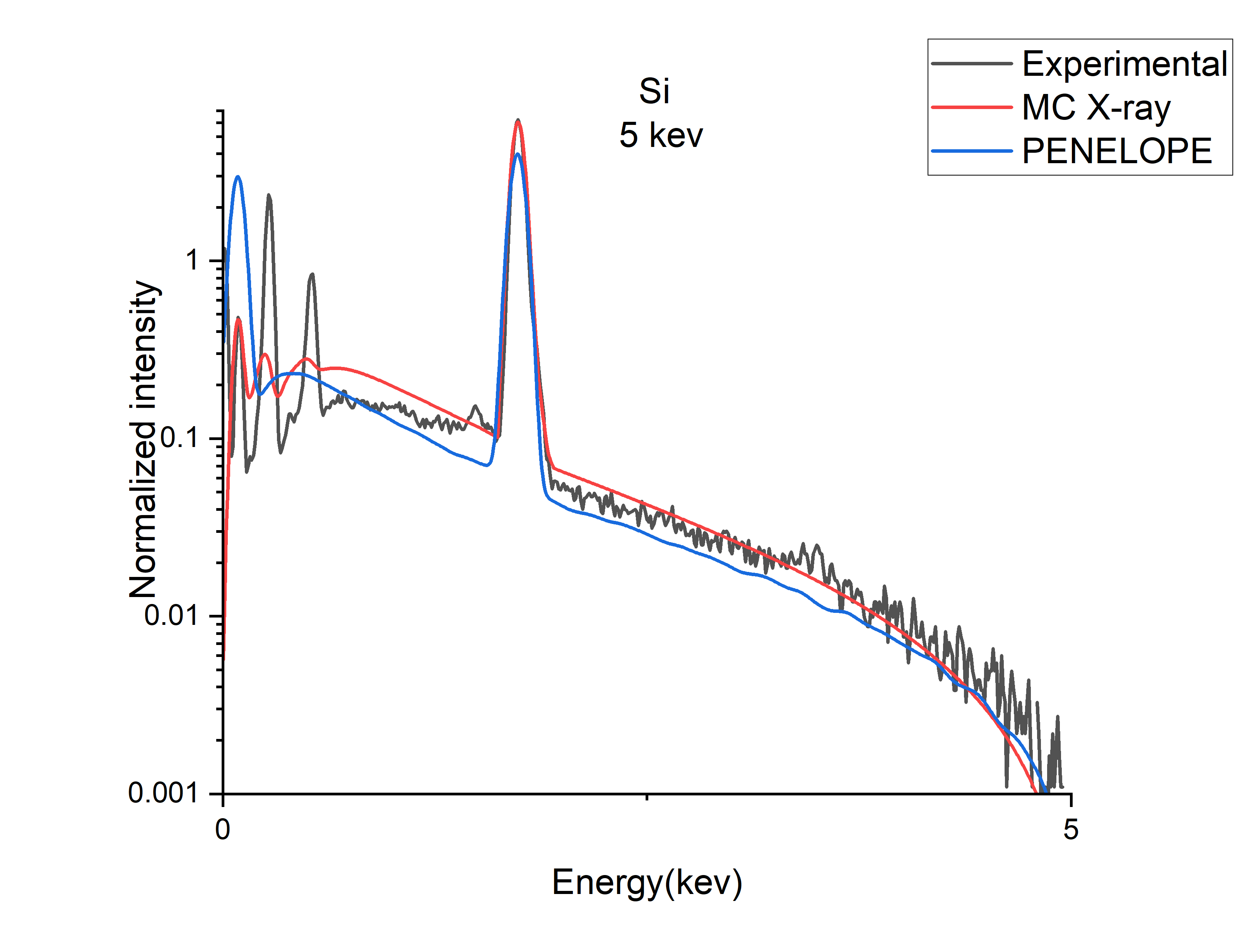}
  \end{minipage}
  \hfill
  \begin{minipage}[b]{0.3\textwidth}
    \includegraphics[width=\textwidth]{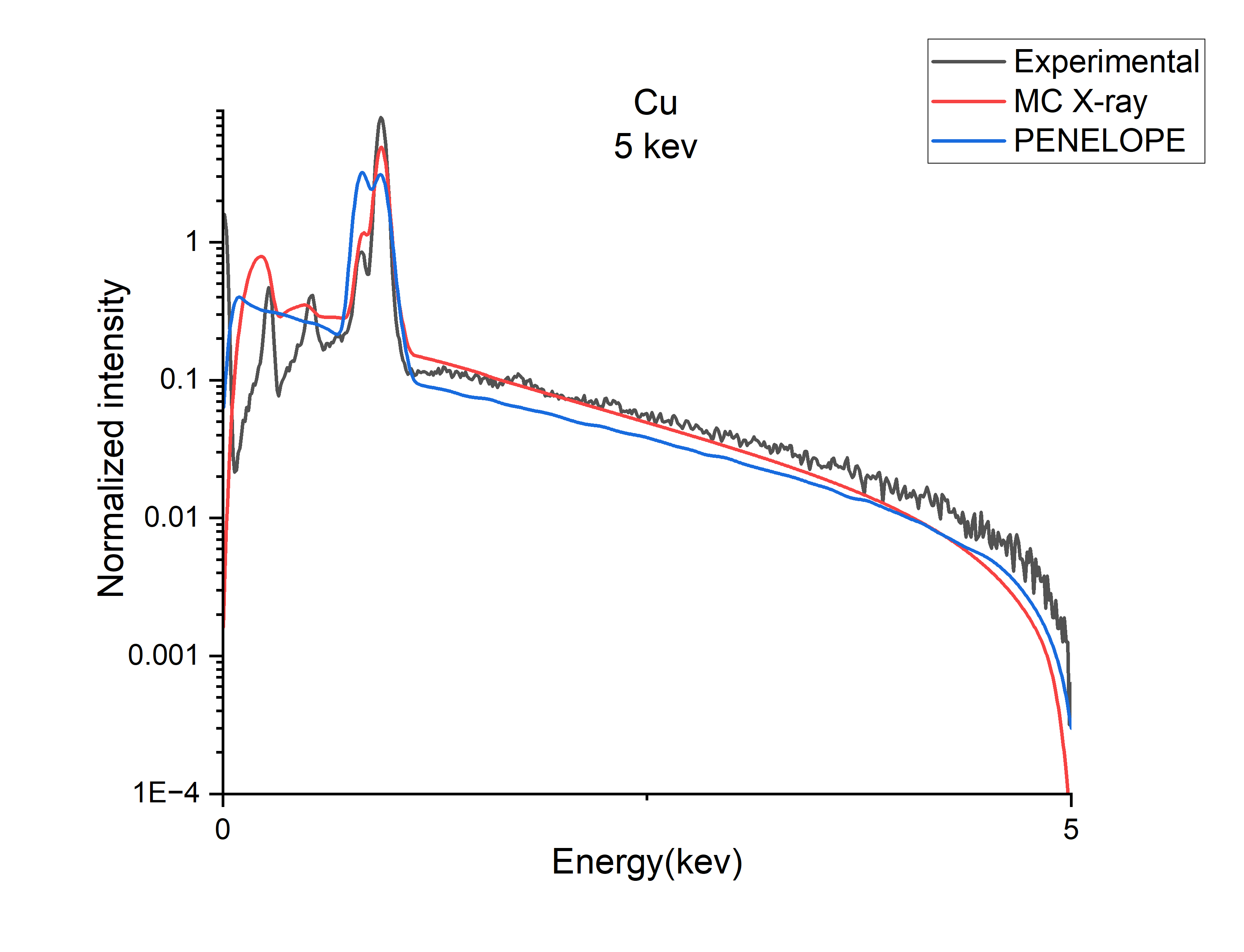}
  \end{minipage}
  \hfill
  \begin{minipage}[b]{0.3\textwidth}
    \includegraphics[width=\textwidth]{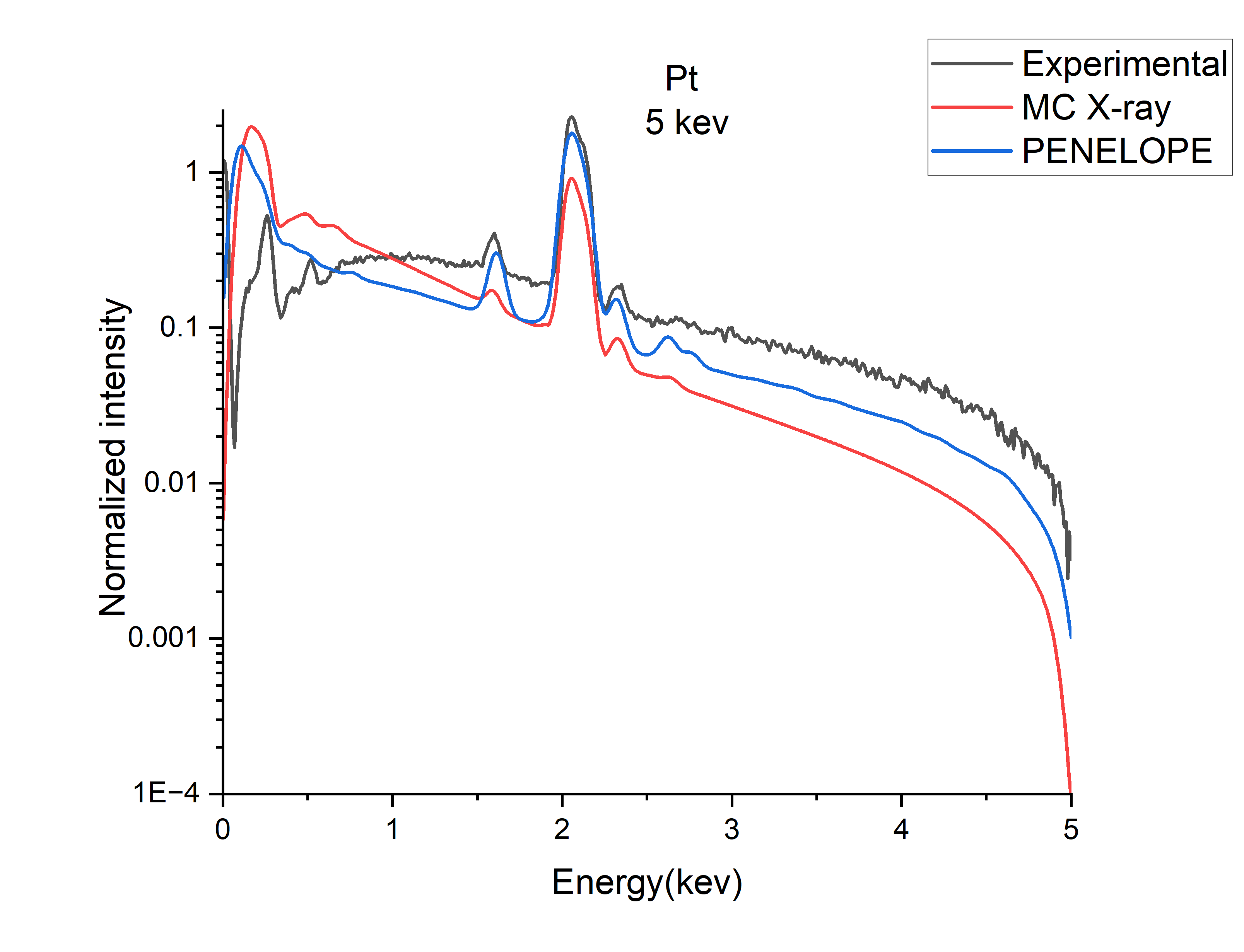}
  \end{minipage}
\end{figure}

\textbf{20 keV}
\begin{figure}[htbp]
  \centering
  \begin{minipage}[b]{0.3\textwidth}
    \includegraphics[width=\textwidth]{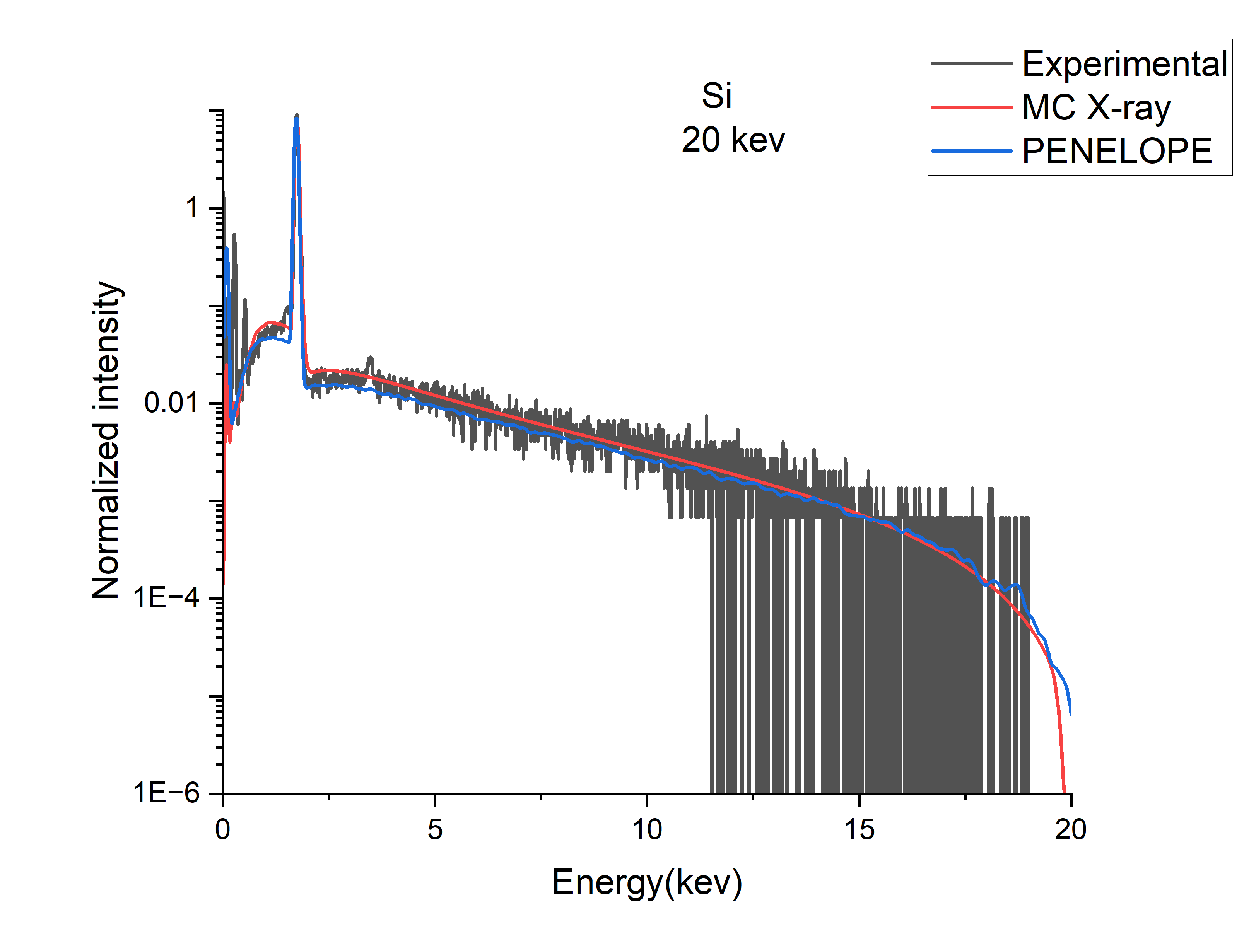}
  \end{minipage}
  \hfill
  \begin{minipage}[b]{0.3\textwidth}
    \includegraphics[width=\textwidth]{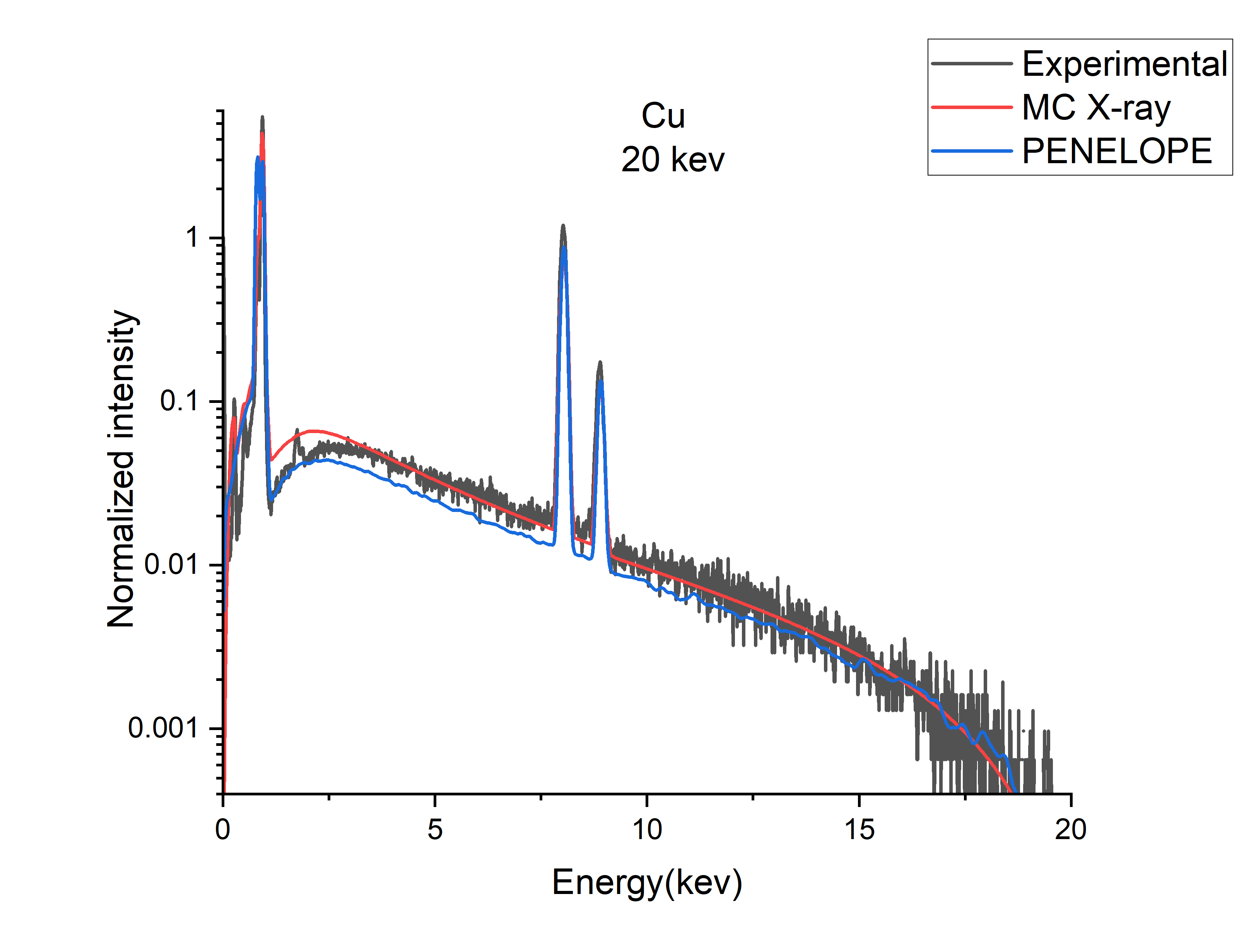}
  \end{minipage}
  \hfill
  \begin{minipage}[b]{0.3\textwidth}
    \includegraphics[width=\textwidth]{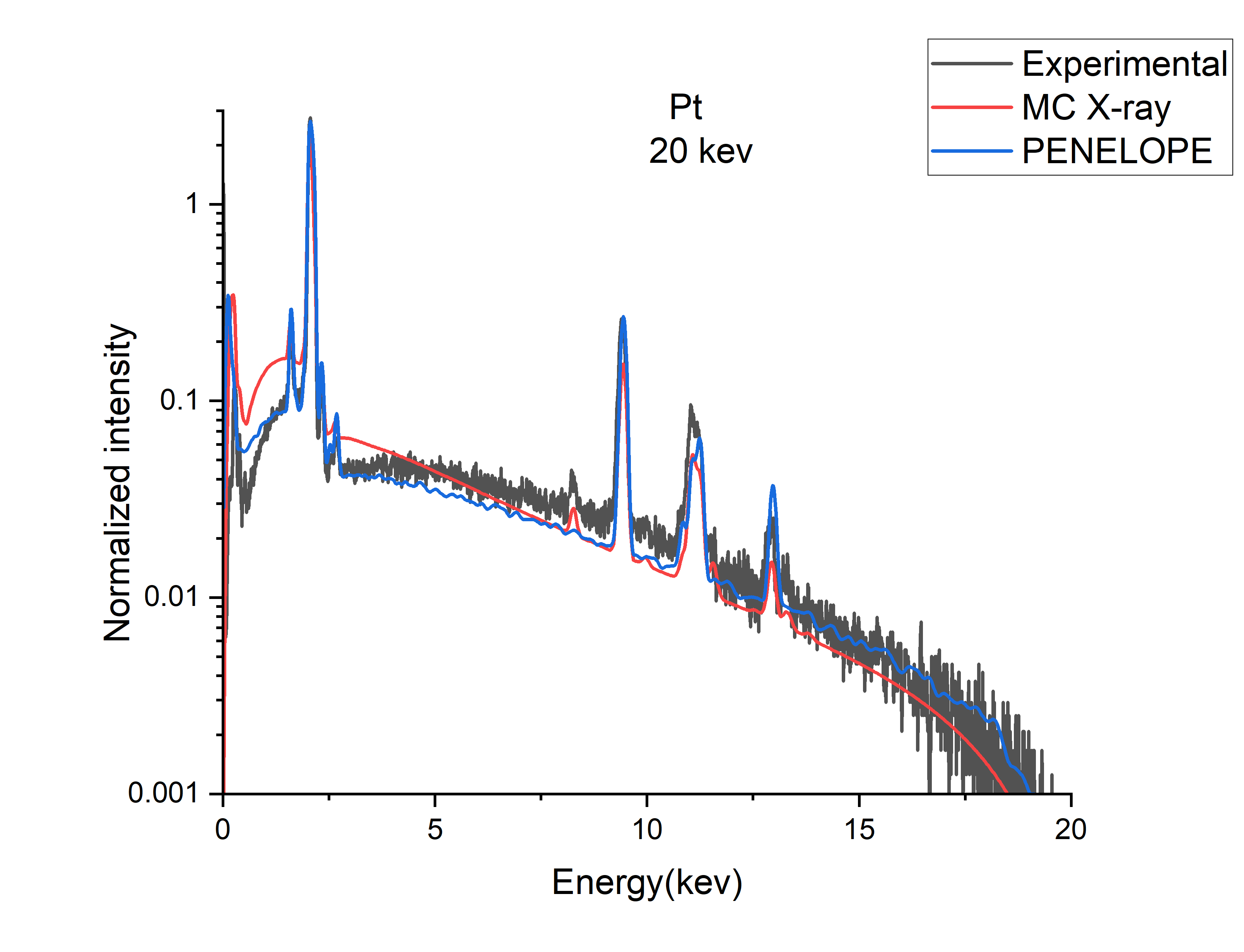}
  \end{minipage}
\end{figure}

\textbf{30 keV}
\begin{figure}[htbp]
  \centering
  \begin{minipage}[b]{0.3\textwidth}
    \includegraphics[width=\textwidth]{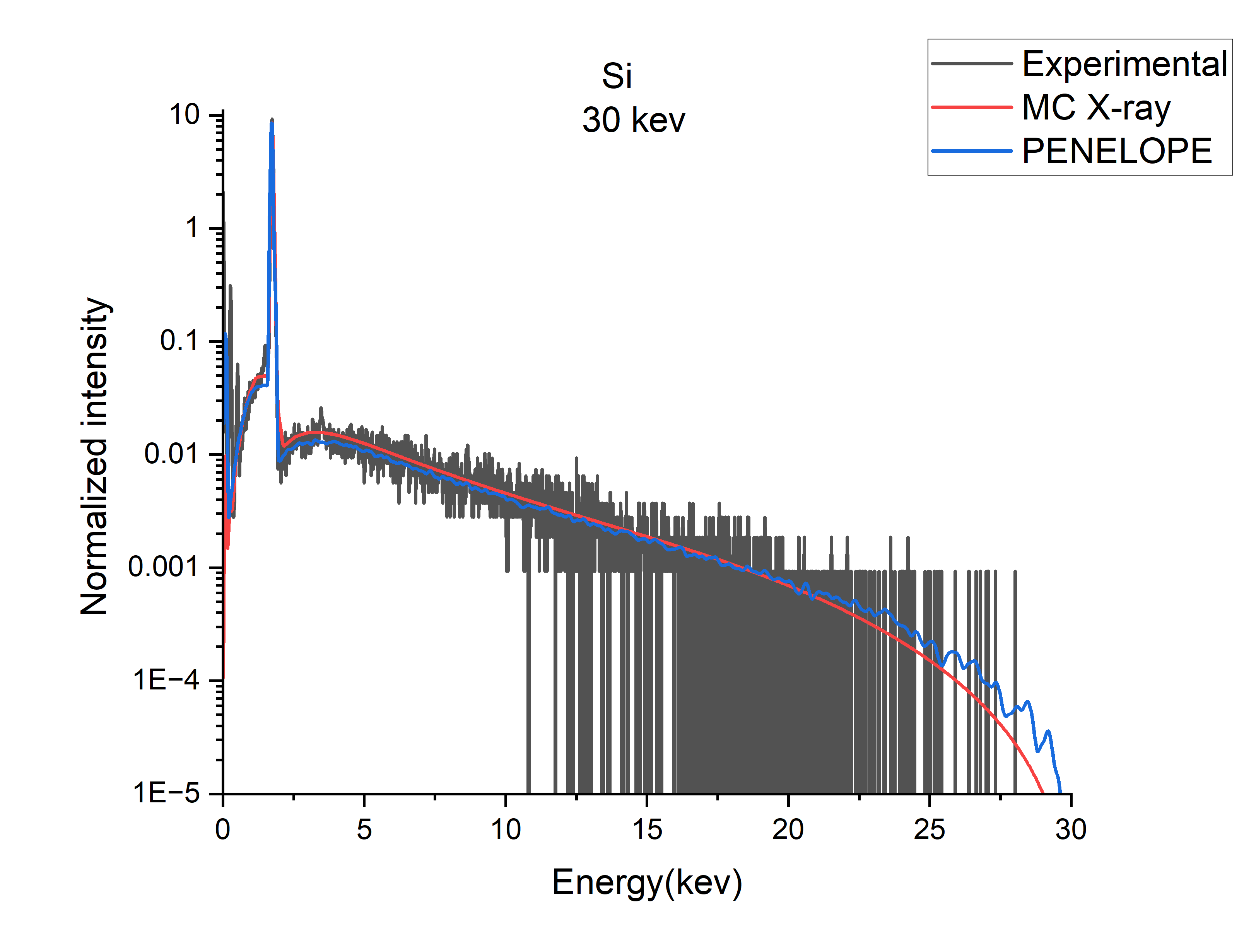}
  \end{minipage}
  \hfill
  \begin{minipage}[b]{0.3\textwidth}
    \includegraphics[width=\textwidth]{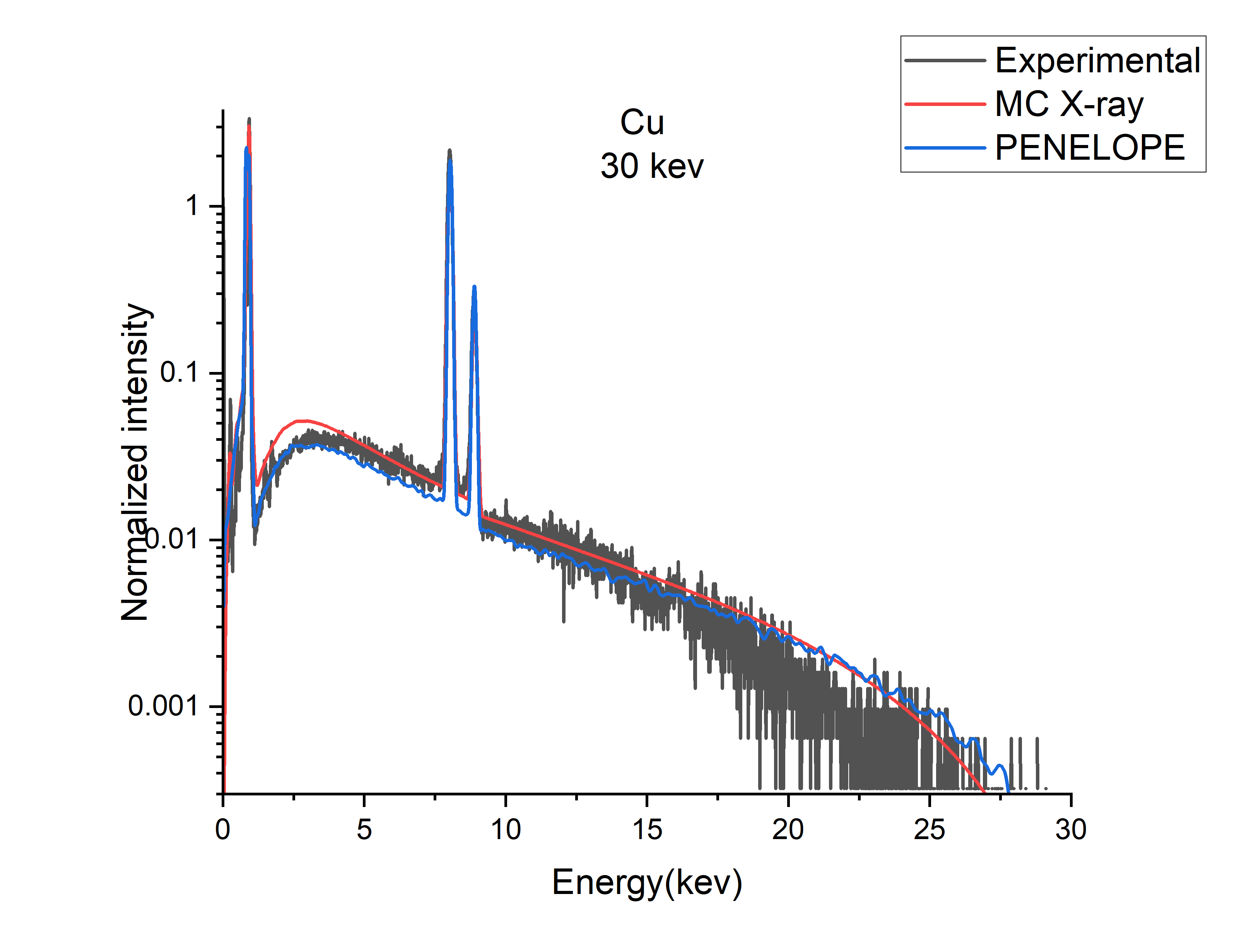}
  \end{minipage}
  \hfill
  \begin{minipage}[b]{0.3\textwidth}
    \includegraphics[width=\textwidth]{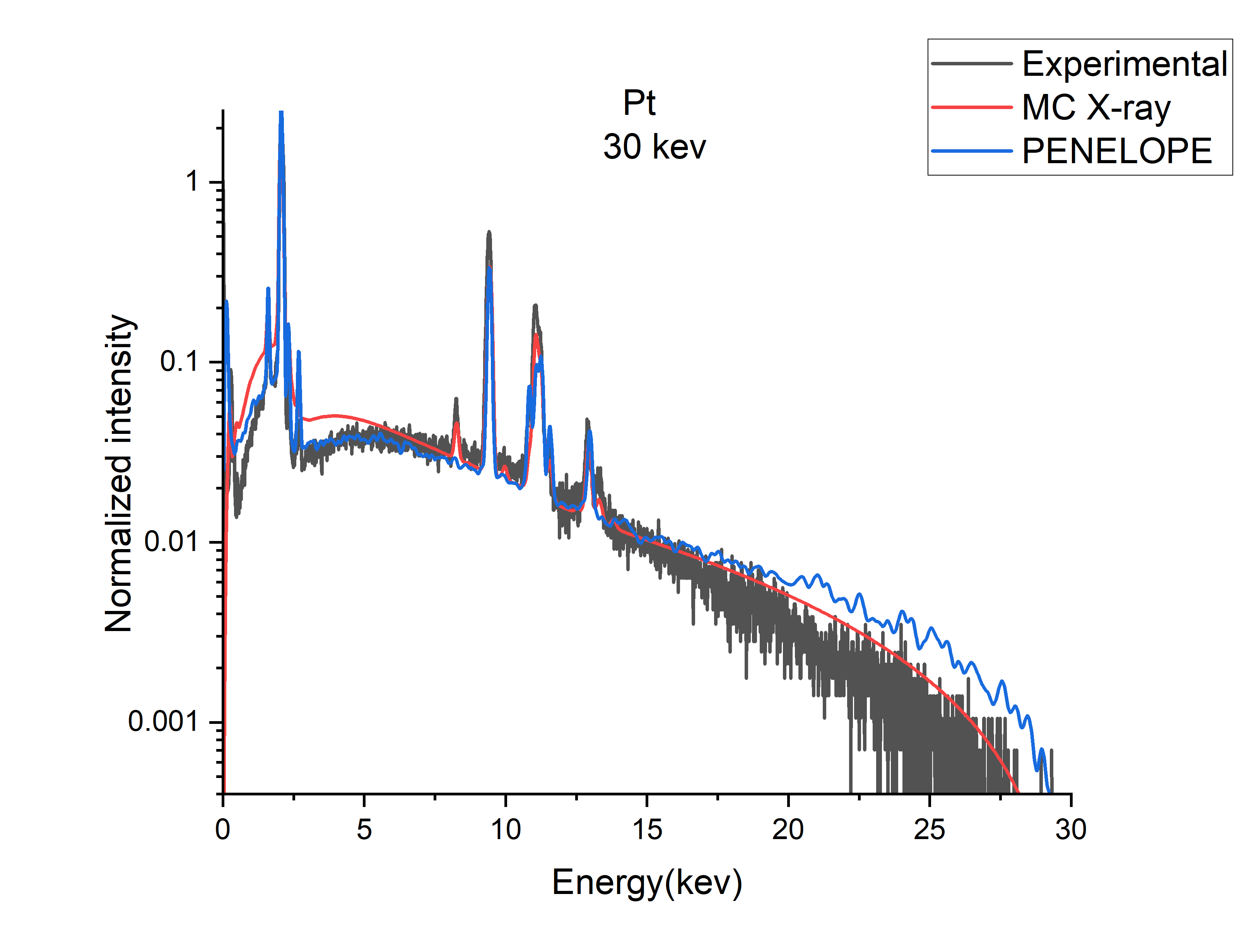}
  \end{minipage}
  \caption{Comparison of Si, Cu, and Pt spectra between MC X-ray, PENELOPE, and experimental data at 3 keV, 5 keV, 20 keV, and 30 keV beam energies. The simulated spectra show notable discrepancies from the experimental data at low beam energies but align closely at higher beam energies.}
\end{figure}

Figure 4 presents the root mean square error (RMSE) values obtained from the comparison between experimental and simulated spectra for 27 materials, evaluated across six specific beam energies. The results show that at relatively low beam energies (3–15 keV), the error increases with the atomic number. However, when the beam energy exceeds 15 keV, the RMSE between the MC X-ray and PENELOPE simulations and the experimental data remains below 0.05. These results indicate that both MC X-ray and PENELOPE generate realistic spectra at beam energies above 15 keV but exhibit significant discrepancies at lower energies.

\begin{figure}[htbp]
  \centering
  \begin{minipage}[b]{0.45\textwidth}
    \includegraphics[width=\textwidth]{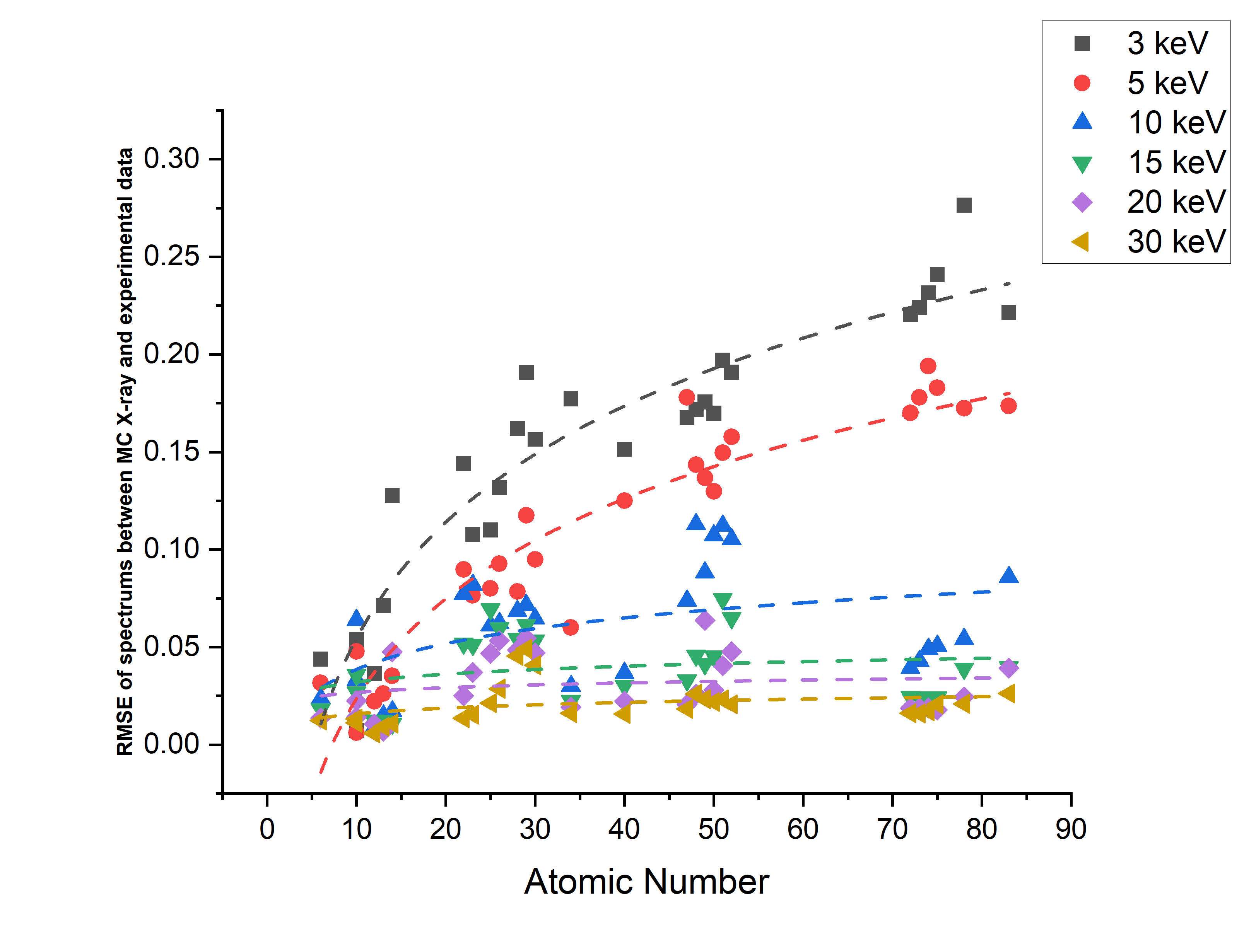}
  \end{minipage}
  \hfill
  \begin{minipage}[b]{0.45\textwidth}
    \includegraphics[width=\textwidth]{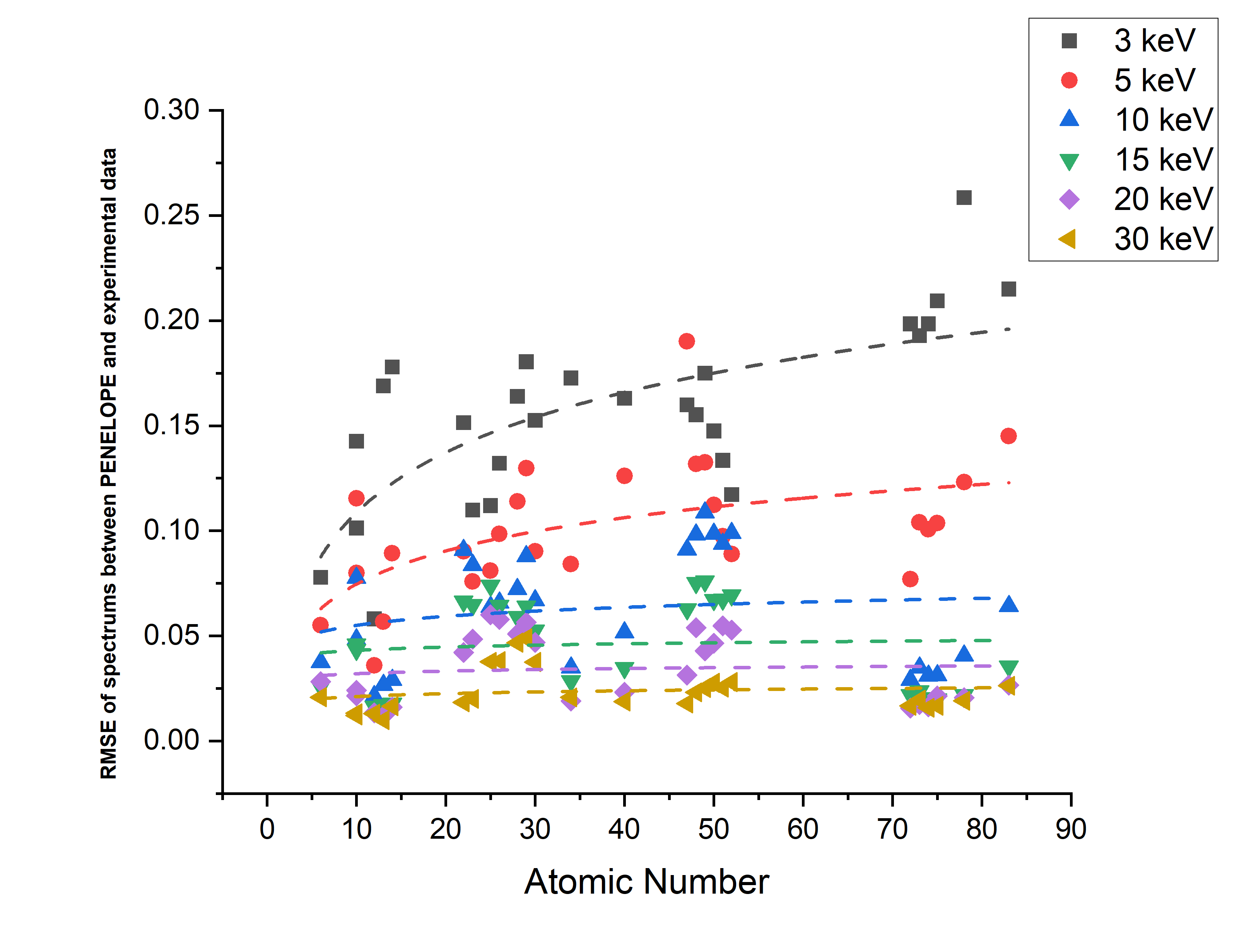}
  \end{minipage}
  \caption{Comparison of simulated spectra from MC X-ray (left) and PENELOPE (right) with experimental data for 27 samples across six different beam energies. Both MC X-ray and PENELOPE produce reliable results, with good alignment between simulated and experimental spectra at higher beam energies}
\end{figure}

\section{Discussion and Conclusion}

The comparison of MC X-ray and PENELOPE in terms of the k-ratio, using the Pouchou database, shows that both software programs achieve similar accuracy. However, MC X-ray demonstrated slightly higher precision, with a root mean square error (RMSE) of 2.71

In terms of efficiency, MC X-ray significantly outperforms PENELOPE. When averaging the simulation time across all 826 samples, MC X-ray required only 50 seconds per simulation, while PENELOPE took 2000 seconds—40 times longer under identical conditions. These simulations were conducted on a personal laptop equipped with an AMD R7 5800H CPU. This substantial time difference highlights MC X-ray’s efficiency, particularly when simulating homogeneous bulk samples with comparable results to PENELOPE.

The discrepancy in simulation time may stem from the different approaches used by the two programs. MC X-ray traces each electron’s trajectory through multiple layers of the sample, with individual electrons providing extensive data on the X-ray distribution. This information is integrated to construct the spectrum and calculate intensities. In contrast, PENELOPE collects all emitted X-rays via detectors to generate the spectrum. Since only a small fraction of incident electrons excite X-rays, PENELOPE requires more electrons, as shown in Figure 3. Even with 40 times longer simulation time, PENELOPE’s bremsstrahlung spectrum displays more fluctuations than MC X-ray’s output. As a result, MC X-ray achieves higher efficiency with fewer electrons.

Additionally, MC X-ray offers greater user convenience. It integrates with Dragonfly, allowing users to employ Python scripts to automate multiple simulations from a single script. By contrast, PENELOPE requires separate, more complex scripts for each simulation. This streamlined workflow, as shown in Tables 3 and 4 of the Appendix, makes MC X-ray more accessible for users seeking efficient and scalable simulation processes.

When comparing the simulated spectra from MC X-ray and PENELOPE with experimental data, some discrepancies arise due to contamination in the experimental samples. Specifically, carbon and oxide contamination contribute to unexpected C and O peaks, especially at low beam energies, where the low signal-to-noise ratio amplifies these discrepancies.

This contamination, along with potential inaccuracies in the bremsstrahlung simulation at low electron energies—particularly at 3 keV and 5 keV—accounts for the observed errors at these lower energies. Figure 4 shows that both MC X-ray and PENELOPE struggle to accurately simulate spectra at such low energies.

Interestingly, the Pouchou database comparison did not show similarly large errors at low beam energies, suggesting that both programs can accurately replicate the characteristic peak areas when contamination is not present. This implies that the main challenges lie in simulating bremsstrahlung X-rays at low energies and in obtaining contamination-free samples, as well as in accurately modeling the thickness of contaminated layers during simulations.

As shown in Figure 3, discrepancies were most evident at low beam energies (3 keV and 5 keV) across three representative samples—Si, Cu, and Pt. However, as the beam energy increased to 20 keV and 30 keV, both MC X-ray and PENELOPE produced reliable spectra. Across 27 samples tested at six different beam energies, both programs demonstrated high accuracy, with RMSE values below 0.05 at higher beam energies (20 keV and 30 keV). These results, combined with MC X-ray’s superior speed and user-friendly interface, suggest that it could generate large volumes of training data for neural networks, enhancing quantitative X-ray microanalysis of real specimens.

In summary, MC X-ray and PENELOPE provide comparable accuracy when simulating homogeneous bulk samples, though minor differences exist in the overall spectral shapes. Both software programs yield realistic intensities, as reflected in the k-ratio comparisons, with MC X-ray achieving slightly better accuracy and precision. Importantly, MC X-ray’s faster simulations and streamlined scripting make it more suitable for handling large-scale workflows. Future studies should explore the performance of both programs on more complex scenarios, such as particles embedded in matrices, thin films, and particles on substrates. Such investigations will provide deeper insights into the strengths and limitations of both programs.

\bibliographystyle{unsrt}

\section{Appendix}

\begin{table}[ht]
\centering
\caption{Part of the Pouchou database}
\begin{tabular}{ccccccc}
\toprule
Analyzed & Line & Companion & Accelerating & Weight & K-ratio & X-ray \\
element  &      & element   & voltage (kV) & fraction & of $A$ & take-off angle \\
\midrule
Al & $K\alpha$ & Fe & 20.0 & 0.2410 & 0.1240 & $52.5^\circ$ \\
Al & $K\alpha$ & Fe & 25.0 & 0.2410 & 0.0980 & $52.5^\circ$ \\
Al & $K\alpha$ & Fe & 30.0 & 0.2410 & 0.0830 & $52.5^\circ$ \\
Fe & $K\alpha$ & Al & 20.0 & 0.7690 & 0.7360 & $52.5^\circ$ \\
Fe & $K\alpha$ & Al & 25.0 & 0.7690 & 0.7420 & $52.5^\circ$ \\
Fe & $K\alpha$ & Al & 30.0 & 0.7590 & 0.7480 & $52.5^\circ$ \\
Fe & $K\alpha$ & S  & 10.0 & 0.4660 & 0.4060 & $75^\circ$ \\
Fe & $K\alpha$ & S  & 12.0 & 0.4660 & 0.4210 & $75^\circ$ \\
Fe & $K\alpha$ & S  & 15.0 & 0.4660 & 0.4250 & $75^\circ$ \\
Fe & $K\alpha$ & S  & 20.0 & 0.4660 & 0.4260 & $75^\circ$ \\
\bottomrule
\end{tabular}
\end{table}

\begin{table}[ht]
\centering
\caption{The list of the comparison elements for the simulated spectrum comparison with real experimental data. For compounds sample, the average atomic number is calculated by (Howell, P. G. T. 1988) $Z = \sum_{i=1}^n C_i Z_i$, where $C_i$ is the atomic fraction, and $Z_i$ is the atomic number for each element.}
\label{tab:elements_comparison}

\noindent
\begin{minipage}{0.5\textwidth}
\centering
\begin{tabular}{lc}
\toprule
Element & Average Atomic Number \\
\midrule
Ag & 47 \\
Al & 13 \\
Al$_2$O$_3$ & 10 \\
Bi & 83 \\
C  & 6  \\
CaCO$_3$ & 10 \\
Cd & 48 \\
Cu & 29 \\
Fe & 26 \\
Hf & 72 \\
In & 49 \\
Mg & 12 \\
Mn & 25 \\
\bottomrule
\end{tabular}
\end{minipage}%
\begin{minipage}{0.5\textwidth}
\centering
\begin{tabular}{lc}
\toprule
Element & Average Atomic Number \\
\midrule
Ni & 28 \\
Pt & 78 \\
Re & 75 \\
Sb & 51 \\
Se & 34 \\
Si & 14 \\
Sn & 50 \\
Ta & 73 \\
Te & 52 \\
Ti & 22 \\
V  & 23 \\
W  & 74 \\
Zn & 30 \\
Zr & 40 \\
\bottomrule
\end{tabular}
\end{minipage}
\end{table}

\lstset{ 
  basicstyle=\ttfamily,
  columns=fullflexible,
  frame=single, 
  breaklines=true, 
}
\begin{table}[H]
\centering 
\caption{Part of the MC X-ray script} 

\begin{lstlisting}
simEnergy = energy * 1000  # in eV
nElectron = 10000

def simulation_settings(contained_elements, compositions):
    elements = contained_elements
    energy = [3, 5, 10, 15, 20, 30]
takeOff = 35

np.save('%s.npy' % (filename), dataEDS)
\end{lstlisting}
\end{table}

\begin{table}[ht]
\centering
\caption{Detailed simulation parameters for the MC X-ray script}

\begin{verbatim}
TITLE  Bulk sample simulation
.
>>>>>> Electron beam definition.
SENERG 30e3                      [Energy of the electron beam, in eV]
SPOSIT 0 0 1                     [Coordinates of the electron source]
SDIREC 180 0                     [Direction angles of the beam axis, in deg]
SAPERT 0                         [Beam aperture, in deg]
.
>>>>>> Material data and simulation parameters.
       Up to 10 materials; 2 lines for each material.
MFNAME Cu.mat                    [Material file, up to 20 chars]
MSIMPA 1e3 1e3 0.1e3 0.2 0.2 1e3 0.1e3  [EABS(1:3),C1,C2,WCC,WCR]
.
>>>>>> Geometry of the sample.
GEOMFN Bulk_geo.geo              [Geometry definition file, 20 chars]
DSMAX  1 1.0e20                  [IB, Maximum step length (cm) in body IB]
.
>>>>>> Interaction forcing.
IFORCE 1 1 4 -5    0.9 1.0       [KB,KPAR,ICOL,FORCER,WLOW,WHIG]
IFORCE 1 1 5 -250  0.9 1.0       [KB,KPAR,ICOL,FORCER,WLOW,WHIG]
IFORCE 1 2 2  10   1e-3 1.0      [KB,KPAR,ICOL,FORCER,WLOW,WHIG]
IFORCE 1 2 3  10   1e-3 1.0      [KB,KPAR,ICOL,FORCER,WLOW,WHIG]
.
>>>>>> Bremsstrahlung splitting.
IBRSPL 1 2                       [KB, splitting factor]
.
>>>>>> X-ray splitting.
IXRSPL 1 2                       [KB, splitting factor]
.
>>>>>> Emerging particles. Energy and angular distributions.
NBE    0.0 30e3 300              [E-interval and no. of energy bins]
NBANGL 90 1                      [Nos. of bins for the angles THETA and PHI]
.
>>>>>> Photon detectors (up to 25 different detectors).
       IPSF=0, do not create a phase-space file.
       IPSF=1, creates a phase-space file.
PDANGL 35 45 0 360 0                   [Angular window, in deg, IPSF]
       .
       >>>>>>>> Job properties
RESUME dump1.dat               [Resume from this dump file, 20 chars]
DUMPTO dump1.dat                  [Generate this dump file, 20 chars]
DUMPP  30                                    [Dumping period, in sec]
       .
RSEED  -10   1                 [Seeds of the random-number generator]
REFLIN 29010300 1 1.5E-3         [IZ*1e6+S1*1e4+S2*1e2,detector,tol.]
NSIMSH 2.0e9                  [Desired number of simulated showers]
TIME   2.0e4                       [Allotted simulation time, in sec]


\end{verbatim}
\end{table}

\end{document}